\documentclass[conference]{IEEEtran}
\usepackage{booktabs} 
\usepackage{graphicx}
\usepackage{amsmath}
\usepackage{amsfonts}
\usepackage{caption}
\usepackage{subfig}
\usepackage{algorithm}
\usepackage[noend]{algpseudocode}
\usepackage{float}
\usepackage{url}
\usepackage{siunitx}
\usepackage{multirow}
\graphicspath{{figs/}} 
\usepackage{etoolbox}
\usepackage{tikz}
\usepackage{slashbox}

\usepackage{times}
\usepackage{xcolor}
\usepackage{soul}
\usepackage[utf8]{inputenc}



\usepackage{algorithm}
\usepackage[noend]{algpseudocode}
\usepackage{xfrac}

\graphicspath{{figs/}} 
\newcommand{\sys}{BlackMarks}

\begin{document}
\title{BlackMarks: Blackbox Multibit Watermarking for Deep Neural Networks}

\author{
    Huili Chen, Bita Darvish Rouhani, and Farinaz Koushanfar\\
    University of California San Diego\\
    huc044@eng.ucsd.edu, bita@ucsd.edu, farinaz@ucsd.edu
} 

\maketitle
\begin{abstract} 

Deep Neural Networks have created a paradigm shift in our ability to comprehend raw data in various important fields ranging from computer vision and natural language processing to intelligence warfare and healthcare. While DNNs are increasingly deployed either in a white-box setting where the model internals are publicly known, or a black-box setting where only the model outputs are known, a practical concern is protecting the models against Intellectual Property (IP) infringement. We propose \sys{}, the first end-to-end multi-bit watermarking framework that is applicable in the black-box scenario. \sys{} takes the pre-trained unmarked model and the owner's binary signature as inputs and outputs the corresponding marked model with a set of watermark keys. To do so, \sys{} first designs a model-dependent encoding scheme that maps all possible classes in the task to bit `0' and bit `1'  by clustering the output activations into two groups. Given the owner's watermark signature (a binary string), a set of key image and label pairs are designed using targeted adversarial attacks. The watermark (WM) is then embedded in the prediction behavior of the target DNN by fine-tuning the model with generated WM key set. To extract the WM, the remote model is queried by the WM key images and the owner's signature is decoded from the corresponding predictions according to the designed encoding scheme. We perform a comprehensive evaluation of \sys{}'s performance on MNIST, CIFAR-10, ImageNet datasets and corroborate its effectiveness and robustness. \sys{} preserves the functionality of the original DNN and incurs negligible WM embedding runtime overhead as low as $2.054\%$. 
\end{abstract}

\section{Introduction} \label{sec:intro}

Deep neural networks and other Deep Learning (DL) variants have revolutionized various critical fields ranging from biomedical diagnosis and autonomous transportation to computer vision and natural language processing~\cite{deng2014deep, ribeiro2015mlaas}. 
Training a highly accurate DNN is a costly process since it requires: (i) processing massive amounts of data acquired for the target application; (ii) allocating substantial computing resources to fine-tune the underlying topology (i.e., type and number of hidden layers), and hyper-parameters (i.e., learning rate, batch size), and DNN weights to obtain the most accurate model. Therefore, developing a high-performance DNN is impractical for the majority of customers with constrained computational capabilities. Given the costly process of designing/training, DNNs are typically considered to be the intellectual property of the model builder and needs to be protected to preserve the owner’s competitive advantage.

Digital watermarking has been immensely leveraged over the past decade for ownership protection in the multimedia domain where the host of the watermark can be images, video content, and functional artifacts such as digital integrated circuits~\cite{furht2004multimedia, hartung1999multimedia, qu2007intellectual}. However, the development of DNN watermarking techniques is still in its early stage. Designing a coherent DNN watermarking scheme for model ownership proof is challenging since the embedded WM is required to yield high detection rates and withstand potential attacks while minimally affecting the original functionality and overhead of the target DNN. 

Existing DNN watermarking techniques can be categorized into two types depending on the application scenario. `White-box' watermarking methods~\cite{uchida2017embedding} assumes the availability of model internals (e.g., weights) for WM extraction while `black-box' watermarking only assumes that the output predictions can be obtained for WM detection~\cite{merrer2017adversarial,yossi}. On the one hand, White-box WMs have a larger capacity (carrying multiple-bit information) but limited applicability due to the strong assumption. On the other hand, black-box WMs enable IP protection for Machine Learning as a Service (MLaaS)~\cite{ribeiro2015mlaas} while only zero-bit watermarking methods have been proposed. It is desirable to develop a systematic watermarking approach that combines the advantages of both types of WMs. While all present black-box watermarking papers embed the WM as a statistical bias in the decision boundaries of the DNN (high accuracy on the WM trigger set), our work is the first to prove that it is feasible to leverage the model's predictions to carry a multi-bit string instead of a one-bit information (existence or not of the WM).

By introducing \sys{}, this paper makes the following contributions:

\begin{itemize}
    \item Proposing \sys{}, the first end-to-end black-box watermarking framework that enables multi-bit WM embedding. \sys{} possesses higher capacity compared to prior works and only requires the predictions of the queried model for WM extraction.  
    
    \item Characterizing the requirements for an effective watermarking methodology in the deep learning domain. Such metrics provide new perspectives for model designers and enable coherent comparison of current and pending DNN IP protection techniques.  
    
    \item Performing extensive evaluation of \sys{}'s performance on various benchmarks. Experimental results show that \sys{} enables robust WM embedding with high detection rates and low false alarm rates. As a side benefit, we find out that \sys{}'s WM embedding improves the robustness of the marked model against adversarial attacks. 
    
\end{itemize}

\begin{figure*}[]
\centering
 \includegraphics[width=0.99\textwidth]{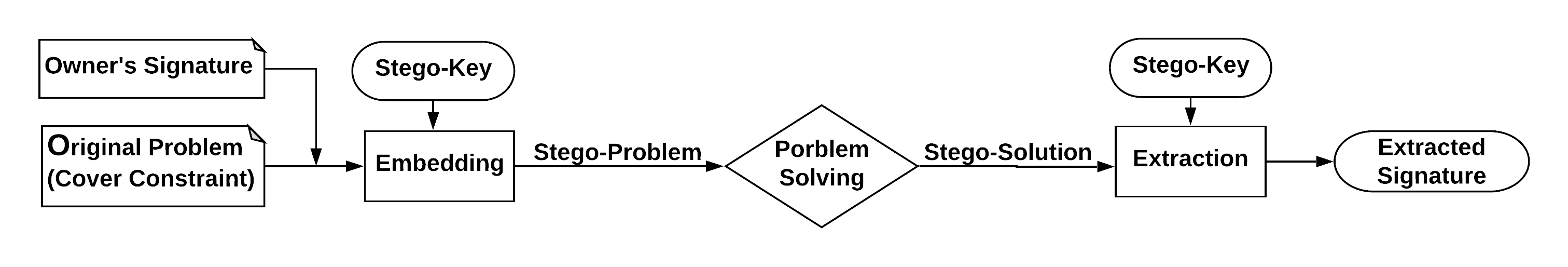}
 \vspace{-1.3em}
\caption{\label{fig:conv_WM} Constraint-based watermarking system. }
\vspace{-0.5em}
\end{figure*}

\section{Preliminaries and Related Work}  \label{sec:related}
\vspace{-0.4em}
Digital watermarks are invisible identifiers embedded as an integral part of the host design and have been widely adopted in the multimedia domain for IP protection~\cite{lu2004multimedia, cox2007digital}. Conventional digital watermarking techniques consist of two phases: WM embedding and WM extraction. Figure~\ref{fig:conv_WM} shows the workflow of a typical constraint-based watermarking system. The original problem is used as the cover constraints to hide the owner's WM signature. To embed the WM, the IP designer creates the stego-key and a set of additional constraints that do not conflict with cover constraints. Combining these two constraints yields the stego-problem, which is solved to produce the stego-solution. Note that the stego-solution will simultaneously satisfy both the original constraints and the WM-specific constraints, thus enables the designer to extract the WM and claim the authorship. An effective watermarking method is required to meet a set of criteria including imperceptibility, robustness, verifiability, capacity, and low overhead~\cite{singh2013survey}.

IP protection of valuable DNN models is a subject of increasing interest to researchers and practitioners. \cite{uchida2017embedding} takes the first step towards DNN watermarking and embeds the WM in the weights of intermediate layers by training the model with an additional regularization loss. The WM is later extracted from the weights at the marked layer assuming a white-box scenario. \cite{rouhani2018deepsigns} present the first generic watermarking approach that is applicable in both white-box and black-box settings by embedding the WM in the activation maps of the intermediate layers and the output layer, respectively. To alleviate the constraint on the availability of model internals during WM extraction, several papers propose zero-bit watermarking techniques that are applicable in the black-box scenario.
\cite{merrer2017adversarial} craft adversarial samples as WMs and embeds them in the decision boundary of the original model by fine-tuning the DNN with the WMs. 
Null hypothesis testing is performed to detect the WM based on the remote model's response to the WM query images. \cite{adi2018turning} suggest to use the incorrectly classified images from the training data as the WM trigger set and generate random labels as the corresponding labels. A commitment scheme is applied to the trigger set to produce the WM marking key and the verification key. The existence of the WM is determined by querying the model with the marking keys and performing statistical hypothesis testing. \cite{zhang2018protecting} proposes three WM generation algorithms (`unrelated', `content', `noise') and embeds the WM by training the model with the concatenation of the training set and the WM set. To detect the WM, the remote model is queried by the WM set and the corresponding accuracy is thresholded to make the decision. To the best of our knowledge, none of the prior works has addressed the problem of multi-bit black-box watermarking.

\vspace{-0.3em}
\section{ \sys{} Overview }  \label{sec:overview}
\vspace{-0.5em}
This section demonstrates the global flow of \sys{} framework (Section~\ref{sec:global}) and introduces a comprehensive set of metrics for an effective DNN watermarking technique (Section~\ref{sec:require}). Potential attacks that might render the embedded WMs undetectable are identified in Section~\ref{sec:attacks}.

\vspace{-0.5em}
\subsection{Global Flow} \label{sec:global}
\begin{figure*}[ht!]
\centering
 \includegraphics[width=0.9\textwidth]{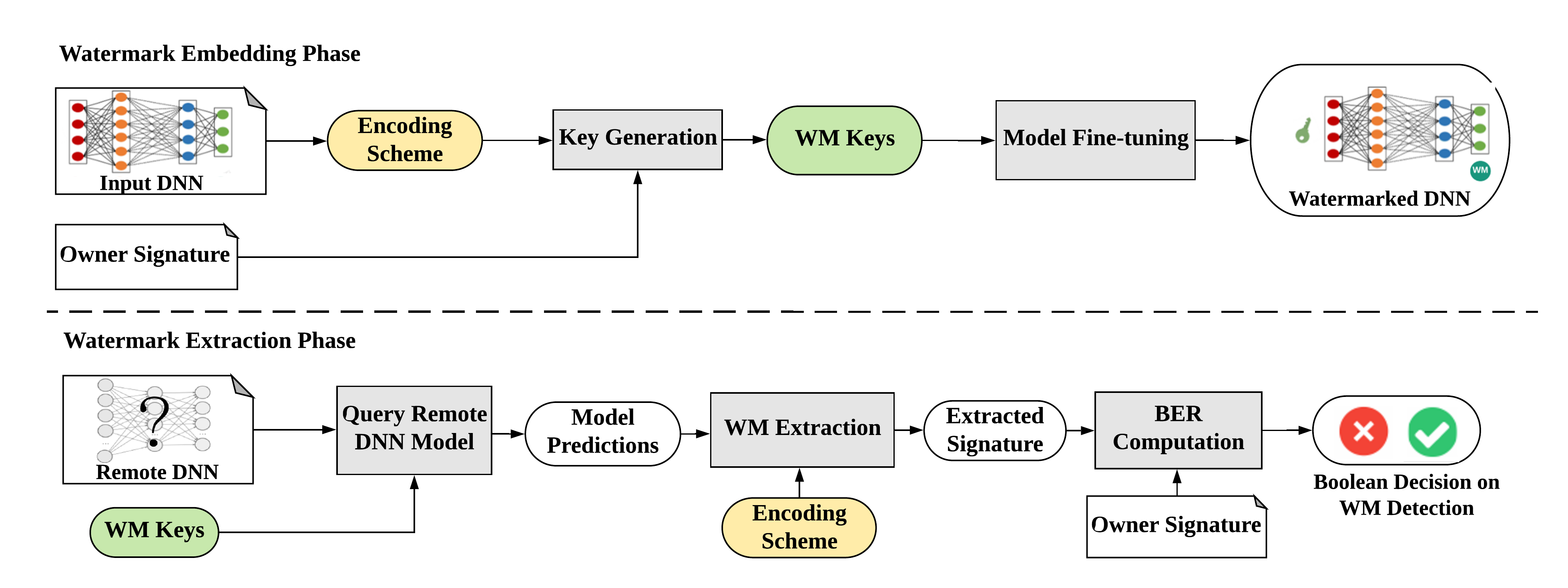} 
 \vspace{-1em}
\caption{\label{fig:global} \sys{} Global Flow: \sys{} watermarking framework has two phases. During WM embedding stage, \sys{} devises a model-specific encoding scheme and generates WM key pairs based on the owner's signature. The designed WM keys are embedded in the target DNN by model fine-tuning. In the WM extraction stage, \sys{} queries the remote DNN with the WM key images and decodes the owner's signature from the model's predictions. }
\end{figure*}

\vspace{-0.3em}
Figure~\ref{fig:global} shows the global flow of \sys{} framework. \sys{} consists of two main phases: watermark embedding and watermark extraction. The marked DNN is deployed as a remote service that only allows API access. \sys{} is the first framework that supports multi-bit DNN watermarking in a black-box setting. We discuss the workflow of each phase as follows.

\begin{table*}
\centering
\caption{Evaluation criteria for an effective watermarking of deep neural networks.}
\label{tab:required}
\scalebox{0.94}{
\begin{tabular}{|l||p{16.2cm}|}
\hline
\multicolumn{1}{|l||}{\textbf{Requirements}}   & \multicolumn{1}{|c|}{\textbf{Description}} \\ \hline \hline
\textbf{Fidelity}     & Accuracy of the target neural network shall not be degraded as a result of watermark embedding. \\ \hline
\textbf{Credibility}  & Watermark extraction shall yield minimal false negatives; WM shall be effectively detected using the pertinent keys. \\ \hline
\textbf{Robustness}   & Embedded watermark shall withstand model modifications such as pruning, fine-tuning, or WM overwriting. \\ \hline
\textbf{Integrity }   & Watermark extraction shall yield minimal false alarms; the authorship of the unmarked models will not be falsely claimed.  \\ \hline
\textbf{Capacity }    & Watermarking methodology shall be capable of embedding a large amount of information in the target DNN. \\ \hline
\textbf{Efficiency}   & Communication and computational overhead of watermark embedding and extraction shall be negligible. \\ \hline
\textbf{Security }    & The watermark shall be secure against brute-force attacks and leave no tangible footprints in the target neural network; thus, an unauthorized party cannot detect/remove the presence of a watermark. \\ \hline
\end{tabular}}
\end{table*}

\noindent \textbf{Watermark Embedding.} The WM embedding module of \sys{} takes the pre-trained model and the owner-specific WM signature as its input. The output is the watermarked DNN together with a set of WM keys. The WM signature is an arbitrary binary string where each bit is independently and identically distributed (i.i.d.). A model-dependent encoding scheme is devised to map all possible labels into bit `0' and bit `1'. The WM embedding process is performed in two steps. First, a set of WM keys are generated as secure parameters for WM embedding. Then, the underlying DNN is trained (fine-tuned) such that the owner-specific WM signature is encoded in the output activations. Note that WM embedding is a one-time task performed by the owner before the model is distributed. Details of each step are discussed in Section~\ref{sec:WM_embed}. 

\noindent \textbf{Watermark Extraction.} To verify the IP of a remote DNN and detect potential IP infringement, the model owner first queries the DNN service with the WM keys generated in the WM embedding phase. \sys{} then decodes the owner's signature from the corresponding triggered predictions using the encoding scheme employed in the embedding stage. The Bit Error Rate (BER) between the extracted signature and the true one is computed. A zero BER implies that the owner's IP is deployed in the remote DNN service. Details of each WM extraction step are discussed in Section~\ref{sec:WM_extract}.

\subsection{Evaluation Criteria}  \label{sec:require}

There are a set of minimal requirements that should be addressed to design a robust digital watermark. Table~\ref{tab:required} details the prerequisites for an effective DNN watermarking methodology. In addition to previously suggested requirements in~\cite{uchida2017embedding, merrer2017adversarial}, we believe credibility and integrity are two other major factors that need to be considered when designing a practical DNN watermarking methodology. \textit{Credibility} is important because the embedded watermark should be accurately extracted using the pertinent keys; the model owner is thereby able to detect any misuse of her model with a high probability. \textit{Integrity} ensures that the IP infringement detection policy yields a minimal false alarms rates, meaning that there is a very low chance of falsely proving the ownership of an unmarked model used by a third party. \sys{} satisfies all the requirements listed in Table~\ref{tab:required} as we empirically show in Section~\ref{sec:eval}.

\subsection{Attack Model} \label{sec:attacks}
\vspace{-0.3em}
To validate the robustness of a potential DL watermarking approach, one should evaluate the robustness of the proposed methodology against (at least) three types of contemporary attacks: (i) \textbf{model fine-tuning}. This attack involves re-training of the original model to alter the model parameters and find a new local minimum while preserving the accuracy. (ii) \textbf{model pruning}. Model pruning is a common approach for efficient DNN execution, particularly on embedded devices. We identify model pruning as another attack approach that might affect the watermark extraction/detection. (iii) \textbf{watermark overwriting}. A third-party user who is aware of the methodology used for DNN watermarking (but does not know the owner's private WM keys) may try to embed a new watermark in the model and overwrite the original one. An overwriting attack intends to insert an additional watermark in the model and render the original one undetectable. A watermarking methodology should be robust against fine-tuning, pruning, and overwriting for effective IP protection.

\section{\sys{} Methodology}
\vspace{-0.5em}
Deep learning models possess non-convex loss surfaces with many local minima that are likely to yield similar accuracy~\cite{choromanska2015loss, rouhani2017deep3}. \sys{} takes advantage of this fact that there is not a unique solution for modern non-convex optimization problems to embed the WM information in the distribution of output activations within the target DNN. We detail the workflow of WM embedding and extraction shown in Figure~\ref{fig:global} in Section~\ref{sec:WM_embed} and~\ref{sec:WM_extract}, respectively. The computation and communication overhead of \sys{} framework is discussed in Section~\ref{sec:overhead}. We use image classification as the host problem in this paper, however, \sys{} can be easily generalized to other data applications.

\subsection{Watermark Embedding}  \label{sec:WM_embed}
\begin{figure}[ht!]
\centering
 \includegraphics[width=\columnwidth]{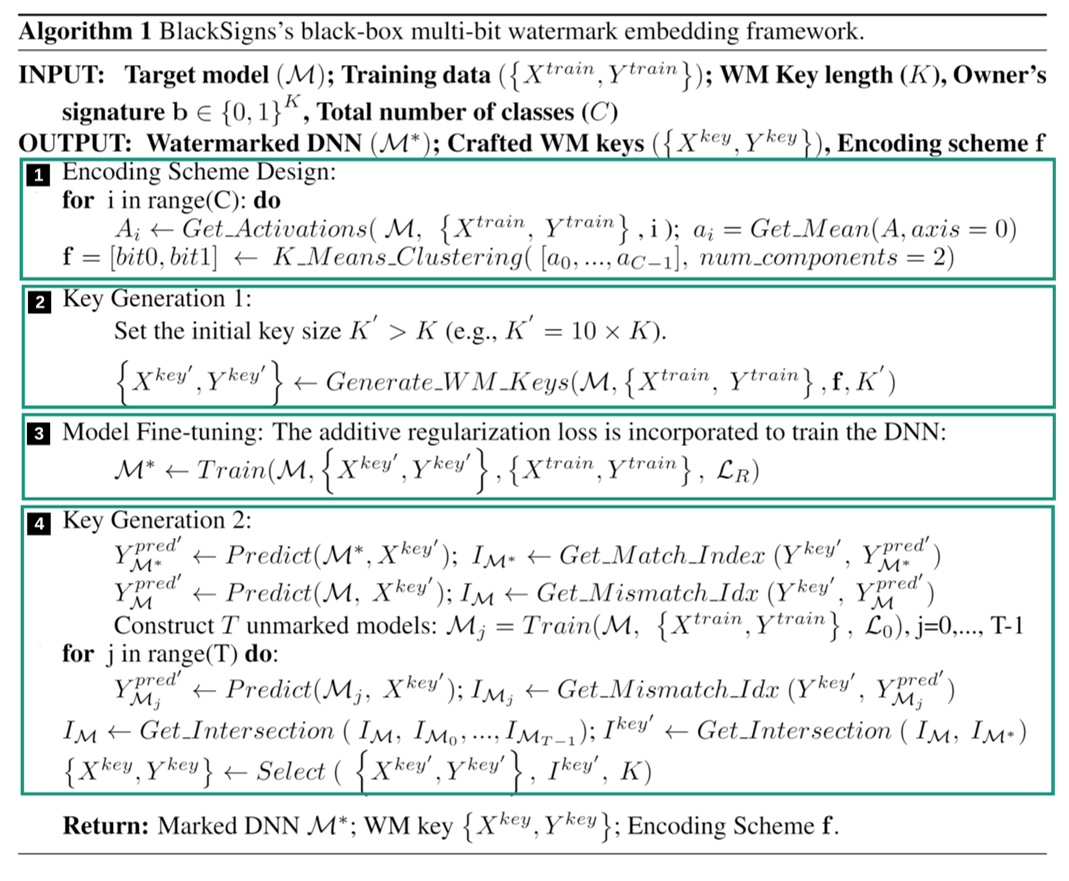} 
 \vspace{-1.2em}
\caption{\label{fig:WM_embed_alg} \sys{}' WM embedding algorithm. }
\vspace{-1.3em}
\end{figure}

\vspace{-0.3em}
Algorithm~1 summarizes the steps of \sys{}' WM embedding. \sys{} encodes the owner-specific WM information in the distribution of the output activations (before \textit{softmax}) while preserving the correct functionality on the original task. The rationale behind our approach is to explore the unused space within the high dimensional DNN~\cite{deepfense} for WM embedding.
\sys{} formulates WM embedding as a one-time, \textit{`post-processing'} step that is performed on the pre-trained DNN locally by the owner before model distribution/deployment. The procedure of \sys{}'s output layer watermarking scheme is summarized in Figure~\ref{fig:WM_embed_alg}. We explicitly discuss each of the steps outlined in Algorithm~1 in the following section. 

\vspace{-1.2em}

\algnewcommand\algorithmicinput{\textbf{INPUT:}}
\algnewcommand\INPUT{\item[\algorithmicinput]}
\algnewcommand\algorithmicoutput{\textbf{OUTPUT:}}
\algnewcommand\OUTPUT{\item[\algorithmicoutput]}

\noindent {\tikz\draw[black,fill=black] (-1em,-1em) rectangle (-0.2em,-0.2em) node[pos=.5, white] {1};} \textbf{Encoding Scheme Design.} Recall that our objective is to encode a binary string (owner's signature) into the predictions made by the DNN when queried by the WM keys. \sys{} designs a model- and dataset-dependent encoding scheme that maps the class predictions to binary bits. The encoding scheme ($\mathbf{f}$) is obtained by clustering the output activations corresponding to all categories ($C$) into two groups based on their similarity. To do so, \sys{} passes a subset of training images in each class through the underlying DNN and acquires the activations at the output layer (before \textit{softmax} applied). The mean value of the output activations in each class is then computed and gathered. Finally, K-means clustering is employed to divide the averaged output activations of all classes into two groups, resulting in the encoding scheme
(step 1). The encoding scheme specifies the collection of labels that correspond to bit `0' and bit `1', respectively.

\noindent {\tikz\draw[black,fill=black] (-1em,-1em) rectangle (-0.2em,-0.2em) node[pos=.5, white] {2};} \textbf{Key Generation 1.} The key generation module takes the encoding scheme obtained in step 1, the owner's private signature ($\mathbf{b} \in \left\{ 0, 1 \right\}^K$, where $K$ is also the key length), and the original training data as inputs and returns a set of image-label pairs as the WM keys for the target DNN. 
To be more detailed, \sys{} deploys \textit{targeted adversarial attacks} to craft WM key images and labels. If the given bit in $\mathbf{b}$ is `0', the source class and the target class for the WM image (`adversarial sample') are determined by uniformly randomly selecting a class label that belongs to the cluster `0' and cluster `1' determined by the encoding scheme ($\mathbf{f}$), respectively. The source class is used as the corresponding WM key label. The WM keys for bit `1' in $\mathbf{b}$ are generated in a similar way.
We use Momentum Iterative Method (MIM)~\cite{dong2018boosting} in our experiments. \sys{} is generic and other targeted adversarial attacks can be used as the replacement of MIM.

\sys{}'s objective is to design specific WM key images as the queries for model authentication instead of crafting standard adversarial samples that cannot be distinguished by human eyes. Therefore, the maximum perturbation is set to $\epsilon=0.5$ with $L_{\infty}$ norm bound. We assume that the WM signature and the key generation parameters (e.g., source and target classes, maximum distortion, step size and the number of attack iterations) are secret information specified by the owner. \sys{}'s WM key images can be considered as a generalization of `adversarial samples' with relaxed constraints on the perturbation level and a different objective (authenticating the ownership of the DNN instead of fooling the model to give wrong results).

It's worth noting that the \textit{transferability} of adversarial samples~\cite{papernot2016transferability, papernot2017practical} might yield false positives of WM detection as shown in~\cite{merrer2017adversarial}. To address this problem, we set the initial key size to be larger than the owner's desired value $K^{'} > K$ and generate the input keys accordingly ($K^{'}= 10 \times K$ in our experiments). The intuition here is that we want to filter out WM keys that are located near the decision boundaries and can easily transfer to unmarked models. 

\noindent {\tikz\draw[black,fill=black] (-1.1em,-1.1em) rectangle (-0.2em,-0.2em) node[pos=.5, white] {3};} \textbf{Model Fine-tuning.} To enable seamless encoding of the WM information, \sys{} incorporates an additive WM-specific embedding loss ($\mathcal{L}_{WM}$) to the conventional cross-entropy loss ($\mathcal{L}_0$) during DNN fine-tuning where a mixture of the WM keys and (a subset of) the original training data is fed to the model. The formulation of the total regularized loss ($\mathcal{L}_R$) is given in Eq.~(\ref{eq:code_loss}). Here, we use \textit{Hamming Distance} as the loss function $\mathcal{L}_{WM}$ to measure the difference between the extracted signature (obtained by $decode\_predictions$) and the true signature $\mathbf{b}$. The embedding strength parameter $\lambda$ controls the contribution of the additive loss term. 

\vspace{-1.3em}
\begin{equation}
    \label{eq:code_loss}
    \mathcal{L}_R = \mathcal{L}_0 +  \lambda \cdot \mathcal{L}_{WM}(~ \mathbf{b}, ~decode\_predictions(Y^{key}_{\mathcal{M}^{*}} ,~\mathbf{f})). 
\end{equation}
\vspace{-1.3em}

Note that without the additional regularization loss ($\mathcal{L}_{WM}$), this retraining procedure resembles `adversarial training'~\cite{kurakin2016adversarial}. All of the existing zero-bit black-box watermarking papers~\cite{merrer2017adversarial, zhang2018protecting, adi2018turning} leverages `adversarial training' for WM embedding to ensure that the marked model has a high classification accuracy on the WM trigger set. However, such an approach does not directly apply to multi-bit WM embedding where we care about the difference between the decoded signature and the original one instead of the difference between the received predictions and the WM key labels. \sys{} identifies this inherent requirement of multi-bit watermarking and formulates an additive embedding loss ($\mathcal{L}_{WM}$) to encode the WM. The rationale behind our design is that, when queried by WM key images, an additional penalty shall be applied if the prediction of the marked model does not belong to the same code-bit cluster as the corresponding WM key label (code-bit clusters are specified by the encoding scheme).

\begin{table*}
\centering
\caption{Benchmark network architectures. Here, ${64C3(1)}$ indicates a convolutional layer with $64$ output channels and ${3\times3}$ filters applied with a stride of 1, ${MP2(1)}$ denotes a max-pooling layer over regions of size ${2\times2}$ and stride of 1, and ${512FC}$ is a fully-connected layer with $512$ output neurons. $BN$ denotes batch-normalization layer. ReLU is used as the activation function for hidden layers. }
\label{tab:bench}
\scalebox{0.92}{
\begin{tabular}{|c||c|c|c|c||c|c|} \hline

\textbf{Dataset} & \textbf{Baseline Accuracy} & \multicolumn{3}{c||}{\textbf{Accuracy of Marked Model}} & \textbf{DL Model Architecture} \\ \hline \hline
\multirow{2}{*}{MNIST} &  \multirow{2}{*}{99.44\%}  & \multicolumn{1}{c|}{K~=~20} & K~=~30 & K~=~50 & 1*28*28-32C3(1)-BN-32C3(1)-MP2(1)-BN \\ \cline{3-5} &   &  \multicolumn{1}{c|}{99.44\%}  & 99.46\% & 99.42\% & -64C3(1)-BN-Flatten-BN-512FC-BN-10FC\\ \hline
                         
\multirow{2}{*}{CIFAR10} &  \multirow{2}{*}{92.09\%} & \multicolumn{1}{c|}{K~=~20}  & K~=~30 & K~=~50  & \multirow{2}{*}{Please refer to \cite{zagoruyko2016wide}.} \\ \cline{3-5}&    &  \multicolumn{1}{c|}{92.10\%} & 92.28\% & 92.19\%  &\\ \hline

\multirow{2}{*}{ImageNet (top-1)} &  \multirow{2}{*}{56.39\%} & \multicolumn{1}{c|}{K~=~20}  & K~=30 & K~=50  & \multirow{2}{*}{Please refer to \cite{krizhevsky2012imagenet}.} \\ \cline{3-5}&    &  \multicolumn{1}{c|}{56.28\%} & 56.31\%  & 56.36\%  &\\ \hline
\end{tabular}
} 
\end{table*}

In our experiments, we set the hyper-parameter $\lambda$ to $0.5$ for MNIST and CIFAR-10 benchmark, and to $0.01$ for ImageNet benchmark. The pre-trained unmarked model is fine-tuned for $15$ epochs with the regularized loss in Eq.~(\ref{eq:code_loss}) for all benchmarks. We use the same optimizer setting used for training the original neural network, except that the learning rate is reduced by a factor of 10. Such retraining procedure coherently encodes the WM key in the distribution of output activations while preventing the accuracy drop on the legitimate data.

\noindent {\tikz\draw[black,fill=black] (-1.1em,-1.1em) rectangle (-0.2em,-0.2em) node[pos=.5, white] {4};} \textbf{Key Generation 2.} Once the model is fine-tuned with the regularized loss in step 3, we first find out the indices of initial WM keys that are correctly classified by the watermarked model (denoted by $I_{\mathcal{M}^{*}}$). To identify and remove WM keys images with high transferability, we construct $T$ variants of the original unmarked model ($T=3$ in our experiments) by regular fine-tuning. All these $(T+1)$ unmarked models are queried to find the common indices of the initial WM keys that are incorrectly classified (denoted by $I_{\mathcal{M}}$). Finally, the intersection of $I_{\mathcal{M}^{*}}$ and $I_{\mathcal{M}}$ (denoted by $I^{key'}$) determines the indices of proper key candidates to carry the WM signature. 
A random subset of candidate WM keys is then selected as the final WM keys according to the owner's key size ($K$). In the global flow (Figure~\ref{fig:global}), we merge the two key generation steps into one module for simplicity. 

\vspace{-0.3em}
\subsection{Watermark Extraction}  
\vspace{-0.3em}
\label{sec:WM_extract} To extract the signature from the remote DNN ($\mathcal{M}^{'}$), the owner queries the model with the WM key images ($X^{key}$) generated in step 4 of WM embedding and obtains the corresponding predictions ($Y^{key}_{M'}$). Each prediction is then decoded to the corresponding binary value using the encoding scheme ($\mathbf{f})$ designed in WM embedding. The decoding is repeated for all predictions on the WM key and yields the recovered signature ($\mathbf{b}^{'}$). Finally, the BER between the true signature ($\mathbf{b}$) and the extracted one ($\mathbf{b}^{'}$) is computed. The owner can prove the authorship of the model if the BER is zero.

\vspace{-0.3em}
\subsection{Watermarking Overhead}  \label{sec:overhead}
\vspace{-0.3em}
Here, we analyze the computation and communication overhead of WM extraction. The runtime overhead of the one-time WM embedding is empirically studied in Section~\ref{sec:eval_overhead}. For the remote DNN service provider, the computation overhead of WM extraction is equal to the cost of one forward pass of WM key images through the underlying DNN. For the model owner, the computation cost consists of two parts: (i) decoding the prediction response $Y^{key}_{M'}$ to a binary vector by finding out which cluster (`0' or `1' in the encoding scheme $\mathbf{f}$) contains each prediction; and (ii) performing an element-wise comparison between the recovered signature ($\mathbf{b}^{'}$) and the true one ($\mathbf{b}$) to compute the BER. In this case, the communication overhead is equal to the key length ($K$) multiplied by the sum of the input image dimension and one to submit the queries and read back the predicted labels.

\section{Evaluation}  \label{sec:eval}
\vspace{-0.5em}
We evaluate the performance of \sys{} framework on various datasets including MNIST~\cite{lecun1998mnist}, CIFAR10~\cite{krizhevsky2009learning} and ImageNet~\cite{imagenet_cvpr09}, with three different neural network architectures. Table~\ref{tab:bench} summarizes DNN topologies used in each benchmark and the corresponding WM embedding results. In Table~\ref{tab:bench}, $K$ denotes the length of the owner's signature (which is also the size of the WM keys). 
In the rest of this section, we explicitly evaluate \sys{}' performance with respect to each requirement listed in Table~\ref{tab:required}. As empirically demonstrated, \sys{} is effective and applicable across various datasets and DNN architectures.


                         


\subsection{Fidelity}  \label{sec:fidelity}
Fidelity requires that the accuracy of the target neural network shall not be significantly degraded after WM embedding. Table~\ref{tab:bench} compares the baseline DNN accuracy (Column 2) and the accuracy of marked models (Column 3 and 4) after WM embedding. As demonstrated, \sys{} respects the fidelity requirement by simultaneously optimizing for the classification accuracy of the underlying model (the cross-entropy loss), as well as the additive WM-specific loss as discussed in Section~\ref{sec:WM_embed}. In some cases (e.g. WideResNet benchmark), we even observe a slight accuracy improvement compared to the baseline. This improvement is mainly due to the fact that the additive loss $\mathcal{L}_{WM}$ in Eq.~(\ref{eq:code_loss}) act as a regularizer during DNN training. Regularization, in turn, helps the model to mitigate over-fitting by inducing a small amount of noise to DNNs~\cite{goodfellow2016deep}.

\begin{table*}[ht!]
\centering
\caption{\sys{}'s robustness against model fine-tuning attacks. The key length is set to $K=50$.}
\label{tab:fine_tuning}
\scalebox{0.9}{
\begin{tabular}{|l|c|c|c|c|c|c|c|c|c|}
\hline
\textbf{Dataset} & \multicolumn{3}{c|}{MNIST} & \multicolumn{3}{c|}{CIFAR-10} & \multicolumn{3}{c|}{ImageNet (top-1)} \\ \hline
\textbf{\# Fine-tuning Epochs} & 20 & 50 & 100 & 20 & 50 & 100 & 5 & 10 & 20 \\ \hline
\textbf{Accuracy} & 99.46\% & 99.48\% & 99.51\% & 92.34\% & 92.36\% & 92.40\% & 55.19\% & 55.13\% & 55.17\% \\ \hline
\textbf{BER} & 0 & 0 & 0 & 0 & 0 & 0 & 0 & 0 & 0 \\ \hline
\textbf{Detection Success} & 1 & 1 & 1 & 1 & 1 & 1 & 1 & 1 & 1 \\ \hline
\end{tabular}
}
\end{table*}

\vspace{-0.3em}
\begin{table*}[ht!]
\centering
\caption{\sys{}'s robustness against watermark overwriting attacks.}
\vspace{-0.3em}
\label{tab:overwrite}
\scalebox{0.9}{
\begin{tabular}{|c|c|c|c|c|c|c|c|c|c|}
\hline
\textbf{Dataset} & \multicolumn{3}{c|}{MNIST} & \multicolumn{3}{c|}{CIFAR-10} & \multicolumn{3}{c|}{ImageNet (top-1)} \\ \hline
\textbf{Key Length} & K=20 & K=30 & K=50 & K=20 & K=30 & K=50 & K=20 & K=30 & K=50 \\ \hline
\textbf{Accuracy} & 99.46\% & 99.43\% & 99.38\% & 92.01\% & 92.09\% & 92.05\% & 56.24\% & 56.32\% & 56.38\% \\ \hline
\textbf{BER} & 0 & 0 & 0 & 0 & 0 & 0 & 0 & 0 & 0 \\ \hline
\textbf{Detection Success} & 1 & 1 & 1 & 1 & 1 & 1 & 1 & 1 & 1 \\ \hline
\end{tabular}
}
\end{table*}

\begin{table*}[ht!]
\centering
\caption{Integrity evaluation of \sys{} framework with various key lengths.}
\label{tab:integrity}
\scalebox{0.87}{
\begin{tabular}{|c|c|c|c|c|c|c|c|c|c|c|c|c|c|c|c|c|c|c|c|}
\hline
\multicolumn{2}{|c|}{\textbf{Dataset}} & \multicolumn{6}{c|}{MNIST} & \multicolumn{6}{c|}{CIFAR-10} & \multicolumn{6}{c|}{ImageNet} \\ \hline
\multicolumn{2}{|c|}{\textbf{Unmarked Models}} & M1 & M2 & M3 & M4 & M5 & M6 & M1 & M2 & M3 & M4 & M5 & M6 & M1 & M2 & M3 & M4 & M5 & M6 \\ \hline
\multirow{3}{*}{\textbf{BER}} & \textbf{K=20} & 0.25 & 0.2 & 0.25 & 0.05 & 0.05 & 0.15 & 0.15 & 0.15 & 0.05 & 0.35 & 0.75 & 0.15 & 0.50 & 0.45 & 0.55 & 0.25 & 0.20 & 0.25 \\ \cline{2-20} 
 & \textbf{K=30} & 0.27 & 0.23 & 0.23 & 0.03 & 0.03 & 0.1 & 0.15 & 0.15 & 0.05 & 0.35 & 0.75 & 0.2 & 0.37 & 0.27 & 0.33 & 0.1 & 0.2 & 0.17 \\ \cline{2-20} 
 & \textbf{K=50} & 0.26 & 0.16 & 0.16 & 0.08 & 0.08 & 0.1 & 0.58 & 0.28 & 0.20 & 0.64 & 0.72 & 0.14 & 0.46 & 0.38 & 0.44 & 0.02 & 0.3 & 0.28 \\ \hline
\end{tabular}
}
\vspace{-1.5em}
\end{table*}

\begin{figure*}[!ht]
 \subfloat{%
  \includegraphics[width=0.32\textwidth]{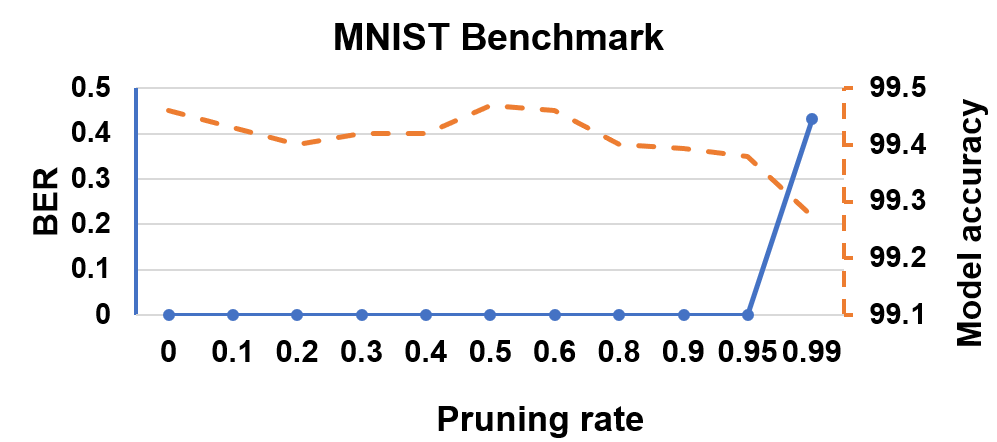}
 }
 \hfill
 \subfloat{%
  \includegraphics[width=0.32\textwidth]{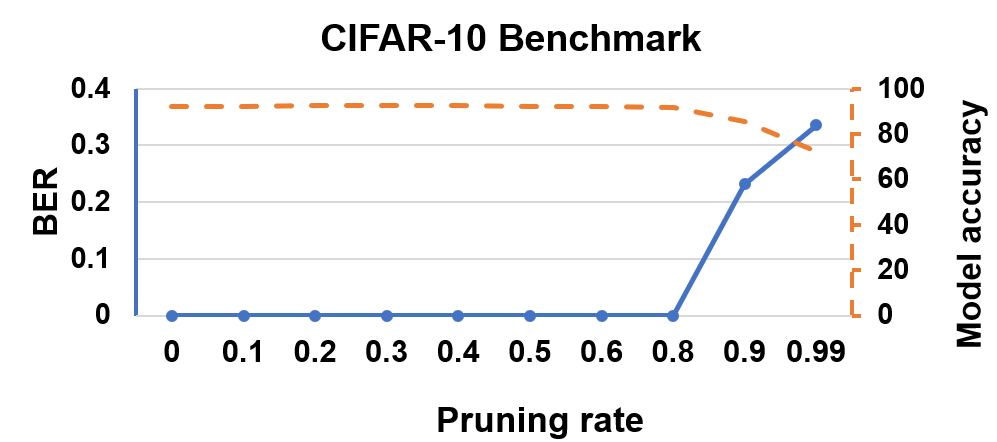}
 }
 \hfill
 \subfloat{
    \includegraphics[width=0.32\textwidth]{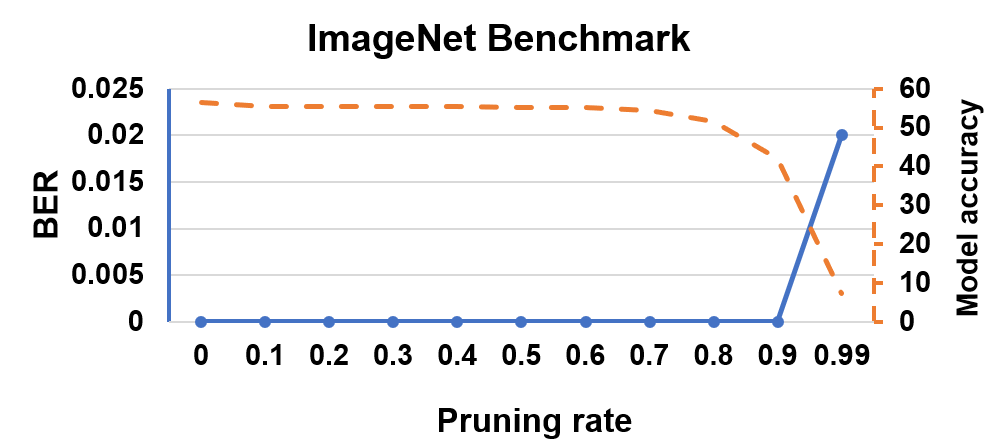}
 }
\vspace{-0.3em}
 \caption{\sys{}' robustness against parameter pruning. The key length is set to $K=50$. The blue solid lines and orange dashed lines denote the BER and final test accuracy, respectively.}
\label{fig:pruning}
\vspace{-0.5em}
\end{figure*}

\subsection{Credibility and Robustness}  \label{sec:robust}
\sys{} enables robust DNN watermarking and reliably extracts the embedded WM for ownership verification. We evaluate the robustness of \sys{} against three state-of-the-art removal attacks as discussed in Section~\ref{sec:attacks}. These attacks include parameter pruning~\cite{han2015learning}, model fine-tuning~\cite{simonyan2014very}, and watermark overwriting~\cite{uchida2017embedding}.

\noindent \textbf{Model Fine-tuning.}
Fine-tuning is a type of transformation attack that a third-party user might use to remove the WM information. To perform such an attack, the adversary retrains the distributed marked model using the original training data with the conventional cross-entropy loss (excluding $\mathcal{L}_{WM}$). Table~\ref{tab:fine_tuning} summarizes the impact of fine-tuning on the watermark detection rate across all benchmarks. As can be seen from the table, the WM signature embedded by \sys{} framework can be successfully extracted with zero BER even after the model is fine-tuned for various numbers of epochs. Note that the number of fine-tuning epochs for ImageNet benchmark is smaller than the other two since the number of epochs needed to train the ImageNet benchmark from scratch is $70$ whereas the other benchmarks take around $200$ epochs to be trained.

\noindent \textbf{Parameter Pruning.} We use the pruning approach proposed in~\cite{han2015learning} to sparsify the weights in the target watermarked DNN. To prune a specific layer, we first set $\alpha\%$ of the parameters that possess the smallest weight values to zero. The model is then sparsely fine-tuned using cross-entropy loss to compensate for the accuracy drop caused by pruning.

Figure~\ref{fig:pruning} demonstrates the impact of pruning on WM extraction. One can see that \sys{} tolerates up to $95\%$, $80\%$, and $90\%$ parameter pruning for MNIST, CIFAR-10, and ImageNet benchmark, respectively. As illustrated in Figure~\ref{fig:pruning}, in cases where DNN pruning yields a substantial BER value, the sparse model suffers from a large accuracy drop. Therefore, one cannot remove \sys{}' WM by excessive pruning while attaining a comparable accuracy with the baseline.

\noindent \textbf{Watermark Overwriting.} Assuming the attacker is aware of the watermarking methodology, he may attempt to corrupt the original WM by embedding a new one. In our experiments, we assume the adversary knows the targeted adversarial attack method employed by the model owner while the owner's signature and the key generation parameters remain secret. In this case, the attacker generates another set of WM keys with his own signature and key generation parameters to fine-tune the marked model following the steps outlined in Algorithm~1. Table~\ref{tab:overwrite} summarizes the accuracy of the overwritten DNN and the BER of the original WM signature ($K=30, K=50$) for all three benchmarks. In our experiments, we assume the attacker uses the same key length as the owner to generate the new WM keys. \sys{} can successfully extract the original WM in the overwritten DNN with zero BER, indicating its credibility and robustness against WM overwriting attacks.

\subsection{Security}   \label{sec:security}

The malicious adversary may try to find the exact WM key designed by the model owner. Since the key generation parameters and the WM signature are assumed to be secret information provided by the owner as discussed in Section~\ref{sec:WM_embed}, even if the attacker is aware of the adversarial attack method used to generate the WM key, he cannot reproduce the exact same key images due to the large searching space. Therefore, \sys{} is secure against brute-force attacks. 
Note that \sys{}'s security derives from the uncertainties in the key generation process.
This is due to the fact that it is sufficient for the attacker to disturb one bit in the recovered WM signature to make the resulting BER non-zero, thus defeating the model authorship proof.

\begin{table*}[ht!]
\centering
\caption{Robustness comparison between unmarked and marked models against adversarial attacks.}
\label{tab:compare_robust}
\scalebox{0.93}{
\begin{tabular}{|c|c|c|c|c|}
\hline
\multirow{2}{*}{\textbf{Accuracy on Adversarial Set}} & \multicolumn{2}{c|}{MNIST} & \multicolumn{2}{c|}{CIFAR-10} \\ \cline{2-5} 
 & Unmarked Model & Marked Model & Unmarked Model & Marked Model \\ \hline
\textbf{FGSM} (untargeted,~\cite{madry2017towards}) & 58.67\% & 75.03\% & 36.51\% & 36.93\% \\ \hline
\textbf{JSMA} (targeted,~\cite{papernot2016limitations}) & 4.98\% & 10.20\% & 76.56\% & 77.21\% \\ \hline
\textbf{MIM} (targeted,~\cite{dong2018boosting}) & 59.24\% & 77.47\% & 53.97\% & 56.82\% \\ \hline
\textbf{CW} (targeted,~\cite{carlini2017towards}) & 87.50\% & 89.91\% & 43.03\% & 44.84\% \\ \hline
\end{tabular}
}
\end{table*}

\begin{figure*}[ht!]
\centering
 \includegraphics[width=0.85\textwidth]{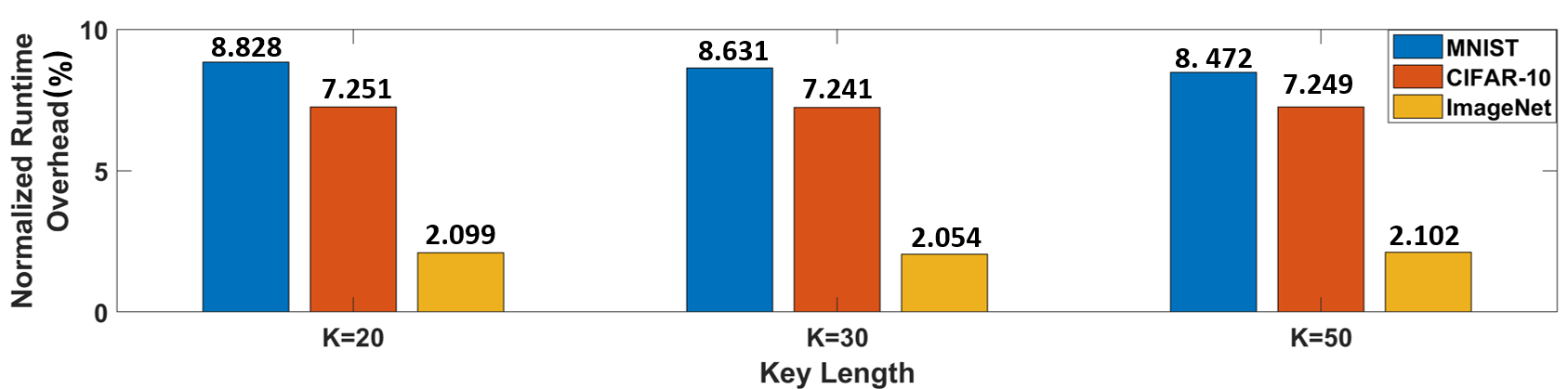} 
 \vspace{-0.2em}
\caption{\label{fig:overhead_WM_embed} Normalized runtime ratio of \sys{}' WM embedding with different key lengths. }
\end{figure*}

\subsection{Integrity} \label{sec:integrity}

Integrity requires that the watermarking technique shall not falsely claim the authorship of unmarked models. For multi-bit watermarking, such requirement means that if an unmarked model is queried by the owner's WM key set, the BER between the decoded signature from the model's predictions and the owner's signature shall not be zero. To evaluate the integrity of \sys{}, we choose six unmarked models for each benchmark and summarize the results in Table~\ref{tab:integrity}. The first three models (M1-M3) have the same network topology but different weights as the watermarked model and the other three models (M4-M6) have different topologies as the marked model. For each benchmark, the owner queries these six unmarked models with her WM keys and tries to extract the WM. The computed BER is \textit{non-zero} in all cases, indicating that \sys{} avoids claiming the ownership of unmarked DNNs and yields low false positive rates.


\subsection{Capacity}  
\label{sec:capacity}
One apparent advantage of \sys{} over existing zero-bit black-box watermarking methods is its higher capacity as we discuss in Section~\ref{sec:related}. To further improve the capacity of the WM, \sys{} can be easily generalized to embed more complex signatures instead of binary vectors. The amount of information carried by the owner's WM signature can be measured by \textit{entropy}~\cite{jaynes1957information}. More generally, if the owner specifies her signature (a numeric vector) with base $B$ and length $K$, the corresponding entropy can be computed as:
\begin{equation}
    \label{eq:entropy_WM}
    H = K \cdot log_2 B
\end{equation}
As can be seen from Eq.~(\ref{eq:entropy_WM}), a longer signature with a larger base value carries more information. Since we use a binary vector ($B=2$) as the WM signature in this paper, the entropy can be simplified as $H=K$. To extend \sys{} framework for embedding a base-$N$ signature, the owner needs to set the number of components in $K\_Means\_Clustering$ to $N$ (Algorithm~1) and change the encoding as well as decoding scheme of predictions correspondingly. 
\sys{} is the first generic multi-bit watermarking framework that possesses high capacity in the black-box setting.

\subsection{Overhead}   \label{sec:eval_overhead}

The WM extraction overhead is discussed in Section~\ref{sec:overhead}. Here, we analyze the runtime overhead incurred by WM embedding. Recall that the WM is inserted in the model by one-time fine-tuning of the target DNN with the regularized loss shown in Eq.~(\ref{eq:code_loss}). As such, the computation overhead to embed a WM is determined by computing the additive loss term $\mathcal{L}_{WM}$ during DNN training. \sys{} has no communication overhead for WM embedding since the embedding process is performed locally by the model owner. To quantify the computation overhead for WM embedding, we measure the normalized runtime time ratio of fine-tuning the pre-trained model with the WM-specific loss and the time of training the original DNN from scratch. To embed the WM, we use the entire training data for MNIST and CIFAR-10 benchmark and $10\%$ of the training data for ImageNet benchmark in our experiments. The selected training data is concatenated with the WM key set to fine-tune the model. The results are visualized in Figure~\ref{fig:overhead_WM_embed}, showing that \sys{} incurs a reasonable additional overhead for WM embedding (as low as 2.054\%) even for large benchmarks.

\section{Discussion}  \label{sec:disc}

Recall that WM embedding leverages a similar approach as `adversarial training' while incorporating a WM-specific regularization loss (Section~\ref{sec:WM_embed}). Here, we study the effect of WM embedding on the model's robustness against adversarial attacks. Table~\ref{tab:compare_robust} compares the robustness of the pre-trained unmarked model and the corresponding watermarked model ($K=50$) against different types of white-box adversarial attacks. It can be seen that for each type of the attack, the marked model has higher accuracy on the adversarial samples compared to the unmarked baseline. The largest accuracy improvement is observed against MIM attacks, which is the method employed by \sys{} to generate WM keys. Such improvement is intuitive to understand since during WM embedding, the first term (cross-entropy loss) in the total regularized loss (see Eq.~(\ref{eq:code_loss})) enforces the model to learn the correct predictions on training data as well as on the WM keys (`adversarial samples'), thus having a similar effect as `adversarial training'~\cite{kurakin2016adversarial}. Therefore, \sys{} has a side benefit of improving the model's robustness against adversarial attacks.

In the future, we plan to extend \sys{} framework to the multi-user setting for fingerprinting purpose. \cite{chen2018deepmarks} present the first collusion-resilient DNN fingerprinting approach for unique user tracking in the white-box setting. 
To the best of our knowledge, no black-box fingerprinting has been proposed due to the lack of black-box multi-bit watermarking schemes. \sys{} proves the feasibility of black-box fingerprinting methods and builds the technical foundation.

\section{Conclusion}  \label{sec:conclusion}

Deep learning is facilitating breakthroughs in various fields and being increasingly commercialized. A practical concern, in the rush to adopt DL models as a service, is protecting the models against IP infringement. In this paper, we propose \sys{}, the first black-box multi-bit watermarking framework for IP protection. To the best of our knowledge, this work provides the first empirical evidence that embedding and extracting multi-bit information using the model's predictions are possible. We introduce a set of requirements to characterize the performance of an effective watermarking technique and perform a comprehensive evaluation of \sys{} accordingly. Experimental results on various datasets and network architectures corroborate that \sys{} coherently embeds robust watermarks in the output predictions of the target DNN with an additional overhead as low as $2.054\%$. \sys{} possesses superior capacity compared to all existing zero-bit watermarking techniques and paves the way for future black-box fingerprinting techniques.

\bibliographystyle{IEEEtran}
\bibliography{IEEEabrv,ref}
\end{document}